\documentclass[english,floatfix,superscriptaddress,nofootinbib,prd]{revtex4-2}
\usepackage[paperwidth=210mm,paperheight=297mm,centering,hmargin=2cm,vmargin=2cm]{geometry}
\usepackage[utf8]{inputenc}
\usepackage[T1]{fontenc}
\usepackage{lmodern}
\usepackage{amsmath}
\usepackage{amssymb}
\usepackage{mathtools}
\usepackage{graphicx}
\usepackage{esint}
\usepackage{dcolumn}
\usepackage{babel}
\usepackage{xcolor}
\usepackage{csquotes}
\usepackage{marvosym}
\usepackage{hyperref}
\usepackage[overload]{textcase}
\usepackage[symbol,hang,flushmargin]{footmisc}

\DefineFNsymbolsTM{otherfnsymbols}{%
  \Biohazard	\circ
  \Bat   \mathsection
  \Yinyang    \dagger
  \Radioactivity \mathsection
}%

\setfnsymbol{otherfnsymbols}

\newenvironment{seqn}{\equation\aligned}{\endaligned\endequation}

\newcommand{\be}{\begin{seqn}}
\newcommand{\ee}{\end{seqn}}

\newcommand{\bea}{\begin{eqnarray}}
\newcommand{\eea}{\end{eqnarray}}
\newcommand{\nn}{\nonumber}

\newenvironment{arabicfootnotes}
  {\par\edef\savedfootnotenumber{\number\value{footnote}}
   
   \setcounter{footnote}{0}}
  {\par\setcounter{footnote}{\savedfootnotenumber}}

\begin{document}
%
%
%
%
%
%
\title{Gravitational Observations and LQGUP}

\author{Mohammed~Hemeda}
\email{mhemeda@sci.asu.edu.eg}
\affiliation{Department of Mathematics, Faculty of Science, Ain Shams University, 11566, Cairo, Egypt.}

\author{Hassan~Alshal}
\email{halshal@scu.edu}
\affiliation{Department of Physics, Santa Clara University, Santa Clara, CA 95053, USA.}
\affiliation{Department of Physics, Cairo University, Giza, 12613, Egypt.}

\author{Ahmed~Farag~Ali}
\email{aali29@essex.edu}
\affiliation{Department of Physics, Faculty of Science, Benha University, Benha, 13518, Egypt.}
\affiliation{Essex County College, 303 University Ave, Newark, NJ 07102, United States .}

\author{Elias~C.~Vagenas}
\email{elias.vagenas@ku.edu.kw}
\affiliation{Department of Physics, College of Science, Kuwait University, Sabah Al Salem University City, P.O. Box 2544, Safat 1320, Kuwait.}
\begin{abstract}
\par\noindent
Motivated by recent works, we employ the bounds on the dimensionless quantum-gravity parameter obtained from six solar system-based gravitational tests in order to obtain bounds on the dimensionless parameter of the generalized uncertainty principle with linear and quadratic terms in momentum. The bounds obtained here are much tighter than those obtained, from the same six solar system-based gravitational tests, for the dimensionless parameter of the uncertainty principle with only quadratic terms in momentum.  
\par\noindent

\end{abstract}
\maketitle
\begin{arabicfootnotes}
%
%
%
%
\section{Introduction}
%
%
%
%
%
\par\noindent
Many theoretical models in the literature nowadays are trying to solve the incompatibility of General Relativity (GR) and Quantum Mechanics (QM) which arises when one tries to combine them in order to get a complete theory of Quantum Gravity (QG) \cite{polchinski_1,polchinski_2,Smolin:2004sx,Rovelli:1990pi,Loll:2019rdj}. One may claim that this combination may occurs by modifying QM without ``touching'' GR, hence by generalizing the main QM principle, i.e., the Heisenberg Uncertainty Principle (HUP), into what is called the Generalized Uncertainty Principle (GUP). The GUP has been considered earlier as a consequence of perturbative string theory by adding an extra quadratic term in momentum after proposing the existence of a minimum measurable length \cite{Amati:1987wq,Gross:1987kza,Amati:1988tn,Konishi:1989wk,Veneziano:1990zh}. Later on, phenomenological approaches were pursued in order to derive a bound on the deformation parameter induced by the GUP \cite{Brau:1999uv,Das:2008kaa,Das:2010sj,Pedram:2011xj,Das:2009hs}.  After this proposal, a series of papers were considered adding an extra linear term in momentum on the top of the quadratic term in momentum,  which gave rises of what is called Linear and Quadratic GUP (LQGUP). This new GUP proposal was motivated by making the uncertainty principle compatible with the Doubly Special Relativity (DSR) theories and consistent with the commutation relations of phase space coordinates $[x_i, x_j ] = [p_i, p_j ] = 0$ via Jacobi identity. The LQGUP is of the following form \cite{Ali:2009zq,Ali:2010yn}
\be
\Delta x \Delta p \geq  \frac{\hbar}{2} \left[1+ \left( \frac{\alpha}{\sqrt{\langle p^2\rangle}} + 4\alpha^2 \right) \Delta p^2 + 4 \alpha^2 \langle p\rangle^2 - 2\alpha \sqrt{\langle p^2\rangle} \right]
\label{lqgup}
\ee
with $\alpha = 2\alpha_0 \ell_p/\hbar$ to be the dimensionful LQGUP parameter, $\ell_p = \sqrt{G\hbar/c^3} = \hbar/2m_p c$ to be the Planck length, and $\alpha_0$ to be the dimensionless LQGUP parameter. Our recent endeavors \cite{Vagenas:2018pez, Vagenas:2019rai, Vagenas:2019wzd, Vagenas:2020bys} motivate us to reassess the value of $\alpha_0$ in light of LQGUP and compare it with that of $\alpha_0$ in light of GUP.

\par\noindent The LQGUP given by Eq. (\ref{lqgup}) can be satisfied by a set of coordinates and momenta defined as
\bea
x_{i} &=& x_{0i}\nn\\
p_{i} &=& p_{0i}\,(1-\alpha p +2 \alpha^{2}p^{2}_{0})
\eea
with $x_{0i}$ and $p_{0i}$ to satisfy the canonical commutation relations   $[x_{0i},p_{0j}]=i\hbar \delta_{ij}$.
\par\noindent
It is well-established \cite{Robertson:1929zz} that an uncertainty principle is produced by an algebra through the inequality of any two observables $\Delta x \Delta p \geq (1/2) |\langle [\hat{x}, \hat{p}]\rangle|$. 
Employing  mirror-symmetric states, i.e., $\langle p\rangle^2 = 0$, one gets $\Delta p = \sqrt{\langle p^2\rangle}$, and the aforesaid LQGUP given in Eq. (\ref{lqgup}) can be written in terms of a commutator as 
\be
\left[x, p \right] = i\hbar \left( 1 - 2\alpha p +4\alpha^2 p^2\right).
\label{commutator}
\ee
The remainder of the paper is structured as follows. In Section II, we present the LQGUP-modified temperature of a Schwarzschild black hole. In Section III, we demonstrate the change in the Schwarzschild metric due to the quantum corrections suggested by LQGUP such that we relate the dimensionless parameter $\alpha_0$ with the corrected term added to the Schwarzschild metric. In Section IV, we reassess the upper bound of the LQGUP parameter $\alpha_0$, using the data obtained from the six solar system-based gravitational tests: light deflection, perihelion precession, pulsar periastron shift, Shapiro time delay, gravitational red shift, and geodetic precession. In Section V, we comment and discuss the obtained results. We compare between the dimensionless parameter of the GUP and that of LQGUP based on the formerly mentioned solar system-based gravitational observations.
%
%
%
%
%
%
\section{LQGUP-modified Black Hole Temperature}
%
%
%
%
\par\noindent
In this section, following the analysis of Ref. \cite{Vagenas:2017vsw}, we will briefly describe the derivation of the LQGUP-modified temperature of a Schwarzschild black hole.
\par\noindent
According to the Heisenberg microscope thought experiment, one needs a photon shot of energy $E$ to locate a particle of size $\delta x$. Within the framework of LQGUP described by Eq. (\ref{lqgup}) and considering the standard dispersion relation for photons, i.e., $E = p$, the size of the particle is
\be
\delta x \sim \frac{\hbar}{2E} - \frac{\hbar \alpha}{2} + 2\hbar \alpha^2 E~.
\label{heisenberg}
\ee
Eq. (\ref{heisenberg}) can be utilized to compute the energy $E$ of the particle by having its (average) wavelength $\lambda \simeq \delta x$. Following the same concept, we can compute the LQGUP-modified  temperature of a Schwarzschild black hole. We consider an ensemble of unpolarized photons as the Hawking radiation particles that are coming out of the event horizon of the Schwarzschild black hole of mass $M$. The event horizon is located at $r_S = 2GM$ (for simplicity we have set $c=k_{B}= 1$). The position uncertainty $\delta x$ of these photons, which is related to the size of the Schwarzschild event horizon $r_S$, will take the form
\be
\delta x \simeq 2\mu r_S = 4\mu GM~,
\label{horizon}
\ee
where $\mu$ is a dimensional parameter that is yet to be determined. According to the equipartition principle, the energy $E$ of the photons of the Hawking radiation is actually the temperature $T$ of the Schwarzschild black hole, i.e., $E = T$. Therefore, from Eqs. (\ref{heisenberg}) and (\ref{horizon}), we obtain the mass-temperature relation as
\be
4\mu GM \simeq \frac{\hbar}{2T} - \frac{\hbar \alpha}{2} + 2\hbar \alpha^2 T~.
\label{mass-temp_1}
\ee
Setting $\alpha=\alpha_{0}/m_{p}$, Eq. (\ref{mass-temp_1}) reads now 
\be
4\mu GM \simeq \frac{\hbar}{2T} - \frac{\hbar \alpha_0}{2m_p} + \frac{2\hbar \alpha_0^2}{m_p^2} T~.
\label{mass-temp_2}
\ee
Considering the semi-classical limit by taking the dimensionless LQGUP  parameter $\alpha_0 \to 0$, the temperature has to be the standard Hawking temperature, namely $T = T_{BH} = \hbar/8\pi GM$, thus we fix the dimensionless parameter to be $\mu = \pi$. Rewriting Eq.(\ref{mass-temp_2}), we get
\be
\left(\frac{4\hbar \alpha_0^2}{m_p^2} \right) T^2 - \left(8\pi GM + \frac{\hbar \alpha_0}{m_p} \right) T + \hbar = 0~.
\label{quad-temp}
\ee
Solving for $T$, we get
\be
T = \frac{\left(8\pi GM + \frac{\hbar \alpha_0}{m_p} \right) \pm \sqrt{\left(8\pi GM + \frac{\hbar \alpha_0}{m_p} \right)^2 -4 \left(\frac{4\hbar \alpha_0^2}{m_p^2} \right) \hbar}}{2 \left(\frac{4\hbar \alpha_0^2}{m_p^2} \right)}~.
\label{temp_1}
\ee
Expanding near the semi-classical limit $\alpha_0 \to 0$, the LQGUP-modified black hole temperature, up to the second order in $\alpha_0$, reads \cite{Vagenas:2017vsw,FaragAli:2015boi}
\be
T = \frac{\hbar}{8\pi GM} \left[1 - \frac{\alpha_0}{2\pi} \left(\frac{m_p}{M} \right) + 5 \left(\frac{\alpha_0}{2\pi} \right)^2 \left(\frac{m_p}{M} \right)^{2} \right]~.
\label{temp_2}
\ee
%
%
%
%
\section{Quantum-Corrected Schwarzschild Metric}
%
%
%
%
\par\noindent
The gravitational interaction between two heavy objects at rest can be described by a potential energy which is produced by the potential generated from the mass $M$ \cite{Donoghue:1993eb,Donoghue:1994dn, Bjerrum-Bohr:2002gqz,Scardigli:2016pjs}
\be
V(r) = -\frac{GM}{r} \left(1 +  \frac{3GM}{r} (1+\frac{m}{M}) + \frac{41}{10\pi} \frac{\ell^2_p}{r^2}\right)~.
\label{potential}
\ee
Due to quantum gravity corrections, the classical Schwarzschild metric can be deformed as
\be
ds^2 = - F(r) dt^2 + F(r)^{-1} dr^2 + r^2 (d\theta^2 + sin^2\theta d\phi^2)
\label{schw-metric}
\ee
with
\be
F(r) = 1 - \frac{2GM}{r} + b(r)~.
\label{g_00}
\ee
Since we are interested only in the leading correction to the Hawking formula, we can consider the simplest deformation of the Schwarzschild metric to be of the form \cite{Scardigli:2014qka}
\be
b(r) = \epsilon\, \frac{G^2 M^2}{r^2}
\label{epsilon}
\ee
\par\noindent
where $\epsilon$ is a dimensionless parameter. This deformation of the Schwarzschild metric is nothing more than the well-known Eddington–Robertson expansion of a spherically symmetric metric. The event horizon of the deformed metric is given by ($F(r_H) =0$)
\be
r_H = r_S \frac{1+\sqrt{1-\epsilon}}{2}
\label{event-horizon}
\ee
with $r_S = 2GM$, and Eq.(\ref{event-horizon}) is valid for $\epsilon \leq 1$. The quantum-corrected Hawking temperature of the deformed metric Eq.(\ref{g_00}) is given by \cite{Scardigli:2014qka,Okcu:2021oke}
\be
T(\epsilon) = \frac{\hbar}{4\pi} F'(r_H) = \frac{\hbar}{\pi r_S} \frac{\sqrt{1-\epsilon}}{(1+\sqrt{1-\epsilon})^2}~.
\label{hawking-temp}
\ee
At this point, we may conjecture that the LQGUP-deformed black hole temperature given by Eq.(\ref{temp_2}) is equal to the quantum-corrected Hawking temperature given by Eq.(\ref{hawking-temp}). Therefore, we obtain
\be
5\left(\frac{1}{2\pi} \frac{m_p}{M} \right)^2 \alpha_0^2 - \left(\frac{1}{2\pi} \frac{m_p}{M} \right) \alpha_0 + X(\epsilon) = 0
\label{quad-alpha_0}
\ee
with
\be
X(\epsilon) = 1- \frac{4\sqrt{1-\epsilon}}{(1+\sqrt{1-\epsilon})^2}~.
\label{x_epsilon}
\ee
Solving for the LQGUP dimensionless parameter $\alpha_0$, we get
\be
\alpha_0 = \frac{1 \pm \sqrt{1-20 X(\epsilon)}}{\frac{5}{\pi} \frac{m_p}{M}}~.
\label{alpha_0}
\ee
\par\noindent
At this point a couple of comments are in order. First, in order to assure that we obtain the semi-classical limit, i.e.,  $\alpha_0 \to 0$ as $\epsilon \to 0$, we need to keep only the negative solution in Eq.(\ref{alpha_0}). Second, if we expand Eq.(\ref{alpha_0}) near the semi-classical limit $\epsilon \to 0$ ($X(\epsilon) \to 0$), we obtain
\be
\alpha_0 = 2\pi \left(\frac{M}{m_p} \right) X(\epsilon)~.
\label{alpha_0-epsilon}
\ee
%
%
%
%
%
%
%
%
\section{LQGUP-Parameter bounds}
%
%
%
%
\par\noindent
In this section, we compute the physical (possible observable) quantities which are modified due to the deformation of the Schwarzschild metric. Consequently, we evaluate the upper bound of LQGUP parameter $\alpha_0$ using the results from six different solar system-based observational tests of GR.
%
%
%
%
%
%
%
%
%
\subsection{Light Deflection}
%
%
%
%
\par\noindent
Here we adopt the analysis of Ref. \cite{Scardigli:2014qka}.  We utilize the polar coordinates $(\phi, r)$ centered at the Sun in order to describe a photon's orbit. The Sun will deflect the orbit of the incoming photon from straight line and, thus, the global deflection angle of the photon's orbit will be given as
\be
\Delta \phi = 2 |\phi(r_0) - \phi(\infty)| - \pi
\label{light-deflection}
\ee
with $r_0$ to be the minimum distance between the photon and the Sun. The photon's orbit from $\infty$ to point $r$ is described by
\be
\phi(r) - \phi(\infty) = \int_\infty^r \frac{dr}{r\sqrt{\left( \frac{r}{r_0} \right)^2 [F(r_0) - F(r)]}}
\label{orbital-change-1}
\ee
where $F(r)$ is given by Eq.(\ref{g_00}). Expanding the above integral in terms of $r_S/r_0$ (with $r_S = 2GM/c^{2}$ to be the unmodified Schwarzschild radius), Eq.(\ref{light-deflection}) will become, to first order in $\epsilon$ and to second order in $r_S/r_0$,
\be
\Delta \phi \simeq 2\left(\frac{r_S}{r_0}\right) + \frac{1}{16} \left(\frac{r_S}{r_0}\right)^2 (15\pi - 16 - 3\pi \epsilon)~.
\label{phi_1}
\ee
\par\noindent
The deflection angle of a photon, that barely touches the surface of the Sun, is normally given by
\be
\Delta \phi = (1+\gamma) \frac{r_S}{r_0}~.
\label{phi_2}
\ee
Comparing Eq.(\ref{phi_1}) with Eq.(\ref{phi_2}), we obtain
\be
|\gamma - 1| = \frac{GM}{8r_{0}c^{2}} \left|15\pi - 16 - 3\pi \epsilon\right|~.
\label{light-defl-gamma-1}
\ee
\par\noindent
Considering the best experimental measurement available for the parameter $\gamma$ from the light bending close to the surface of the Sun, we obtain \cite{Will:2014kxa,Shapiro:2004zz,Lambert:2009xy}
\be
\frac{GM}{8r_{0}c^{2}} |15\pi - 16 - 3\pi \epsilon| \lesssim 1.6 \times 10^{-4}~.
\label{light-defl-obs}
\ee
\par\noindent
Then, taking $M = M_\odot=1.989\times 10^{30}$~kg to be the solar mass, $r_0 = R_\odot=6.963\times 10^{8}$~m to be the radius of the Sun, and implementing the mathematical  constraint $\epsilon \leq 1$, we obtain\footnote{The Newton's constant is $G=6.674 \times 10^{-11}$~Nm$^{2}$/kg$^{2}$ and the speed of the light is $c=3\times 10^{8}$~m/s.}
\be
-65 \lesssim \epsilon \leq 1~.
\label{light-defl-epsilon}
\ee
\par\noindent
Employing the lower bound of Eq.(\ref{light-defl-epsilon}) in Eq.(\ref{alpha_0-epsilon}), we obtain the upper bound of the dimensionless LQGUP parameter, i.e., $\alpha_0$, to be
\be
|\alpha_0| \leq 3.50 \times 10^{38}~.
\label{light-defl-alpha_0}
\ee
It is evident that this bound is not a strict one among the bounds obtained from the solar system-based gravitational tests, and we will see that tighter bounds will be obtained in the coming subsections.
\subsection{Perihelion Precession}
%
%
%
%
\par\noindent
We adopt the analysis of Ref. \cite{Scardigli:2014qka} and use polar coordinates to study a planet orbiting around the Sun. As expected, the planet's orbit  is an elliptical one.  The total orbital precession in each revolution is 
\be
\Delta \phi = 2 |\phi(r_+) - \phi(r_-)| -2\pi
\label{preihelion-prec}
\ee
with $r_{+}$ to be the maximum distance of the planet from the Sun, namely the aphelion, and $r_{-}$ to be the minimum distance of the planet from the Sun, namely the perihelion. For an arbitrary point $r$, the orbital precession as the planet moves from $r_-$ to $r$ is given by the integral
\be
\phi(r) - \phi(r_-) = \int_{r_-}^r \frac{dr}{r^2 \sqrt{F(r) \left[ \frac{r_-^2 \left( \frac{1}{F(r)} - \frac{1}{F(r_-)}\right) - r_+^2 \left( \frac{1}{F(r)} - \frac{1}{F(r_+)}\right)}{r_-^2 r_+^2  \left( \frac{1}{F(r_+)} - \frac{1}{F(r_-)}\right)} - \frac{1}{r^2} \right]}}
\label{orbital-change-2}
\ee
\par\noindent
where $F(r)$ is given by Eq.(\ref{g_00}). Expanding the above integral in terms of $r_S/L$ (with $r_S = 2GM$ to be the unmodified Schwarzschild radius, $L$ to be the {\it semilatus rectum}, and here, for simplicity, $c=1$), the total precession after one revolution is given as, to second order in $\epsilon$ and $r_S/r_0$,
\be
\Delta \phi \simeq \pi \left(\frac{6 - \epsilon}{2}  \right) \left( \frac{r_S}{L} \right) + \frac{\pi}{2} \left(\frac{r_S}{L}  \right)^2 N(\epsilon, e)
\label{preihelion-phi-1}
\ee
or, equivalently,
\be
\Delta \phi \simeq 2\pi \left(\frac{6 - \epsilon}{2}  \right) \left( \frac{GM}{L} \right) + 2\pi \left(\frac{GM}{L}  \right)^2 N(\epsilon, e)
\label{preihelion-phi-2}
\ee
where
\be
N(\epsilon, e) = \frac{1}{2} \left[19 - 8\epsilon + (3-\epsilon) \frac{e^2}{2} - \frac{\epsilon^2}{4}  \right]
\label{n_epsilon}
\ee
with $e$ to be the eccentricity. 
Next, we write Eq.(\ref{preihelion-phi-2}) to first order in $r_S/L$ so that we get
\be
\Delta \phi \simeq \frac{6 \pi GM}{L} \left(1 - \frac{\epsilon}{6}  \right)
\label{preihelion-phi-3}
\ee
which leads to the GR prediction as $\epsilon \to 0$. 
Now, we can compare Eq.(\ref{preihelion-phi-3}) with the best known and measured precession in the solar system, the perihelion precession for Mercury \cite{Will:2014kxa}. The latest data is given by
\be
\Delta \phi_{obs} = \frac{6\pi GM}{L} \left[\frac{1}{3} (2 + 2\gamma - \overline{\beta}) + 3 \times 10^{-4} \frac{J_2}{10^{-7}} \right]
\label{preihelion-phi-obs}
\ee
\par\noindent
where $\Delta \phi_{obs}$ is the observed perihelion shift, $J_2$ is a dimensionless measure of the quadrupole moment of the Sun, and $\gamma$ and $\overline{\beta}$ are the usual Eddington-Robertson expansion parameters. Comparing Eq.(\ref{preihelion-phi-obs}) with Eq.(\ref{preihelion-phi-3}), we get
\be
|\epsilon| \lesssim 6 \times 10^{-3}
\label{preihelion-epsilon}
\ee
where, from observational data for the Sun and Mercury \cite{Weisberg:2010zz}, the bound $|2\gamma - \overline{\beta} -1| \lesssim 3 \times 10^{-3}$ has been implemented, and the last term in Eq.(\ref{preihelion-phi-obs}) has been neglected since it is smaller than the observational error. Substituting the above bound on $\epsilon$ into Eq.(\ref{alpha_0-epsilon}), we obtain the upper bound of the dimensionless LQGUP parameter, i.e., $\alpha_0$, to be
\be
|\alpha_0| \lesssim 1.28 \times 10^{33}
\label{preihelion-alpha_0}
\ee
\par\noindent
which is much tighter bound than the one obtained using the light deflection observational test.
%
%
%
%
%
%
\subsection{Pulsar Periastron shift}
%
%
\par\noindent
We adopt the analysis of Ref. \cite{Scardigli:2014qka}, and we compare Eq.(\ref{preihelion-phi-3}) with another very good observational measurement of the pulsars periastron shift. The observed value of the periastron shift is $\Delta \phi_{obs} \simeq 4.226598(5)$ \cite{Weisberg:2010zz}, where the number in parentheses represents the uncertainty in the last quoted digit. Therefore, the relative error with respect to the GR theoretical prediction, namely $\Delta \phi_{GR}$, of the periastron shift can be defined as
\be
\tilde \epsilon = \frac{\Delta \phi_{obs} - \Delta \phi_{GR}}{\Delta \phi_{GR}}
\label{periastron-shift-error}
\ee
which may be written as $\Delta \phi_{obs}=\Delta \phi_{GR} (1+\tilde \epsilon)$ and can be compared with $\Delta \phi$ in Eq.(\ref{preihelion-phi-3}) to get $|\epsilon| = 6 |\tilde \epsilon|$. Using some numerical measurements of periastron shift \cite{Weisberg:2010zz}, we get $|\tilde \epsilon| = 8.9 \times 10^{-5}$, which gives
\be
|\epsilon| \simeq 5.4 \times 10^{-4}~.
\label{periastron-epsilon}
\ee
\par\noindent
Considering this bound on the dimensionless parameter of the deformed metric and employing it in Eq. (\ref{alpha_0-epsilon}), we obtain the upper bound of the dimensionless LQGUP parameter, i.e., $\alpha_0$, to be
\be
|\alpha_0| \lesssim 2.96 \times 10^{31}
\label{periastron-alpha_0}
\ee
which is the strictest bound among the bounds obtained in this work from the solar system-based gravitational tests. 
%
%
%
%
\subsection{Shapiro Time Delay}
%
%
%
\par\noindent
Considering Irwin Shapiro observational endeavors on measuring the time delay of photon due to gravitational field \cite{Shapiro:1964uw}, we study the impact of the LQGUP-deformed
Schwarzschild metric on calculating the delay in the traveling time of electromagnetic signal moving from point $A$ at $r_A$ to point $B$ at $r_B$  and reflected back to A due to gravitational field of the solar system. We adopt the analysis of Ref. \cite{Okcu:2021oke}, and we obtain
\be
\delta t = 4GM \left( 1 + \ln \left( \frac{4r_A r_B}{r_C^2} \right) - \frac{2GM}{r_C} \left(1 + \frac{3(\epsilon-5)\pi}{8} \right) \right)
\label{time-delay}
\ee
with $r_C$ to be the position of the closest point $C$ the electromagnetic signal would pass by near the Sun surface as the Sun is located at the center of the system under study.
\par\noindent
The time delay in Parameterized Post-Newtonian (PPN) formalism is \cite{Will:2014kxa} 
\be 
\delta t = 4GM \left(1 + \left(\frac{1+\gamma}{2} \right)  \ln \left(\frac{4r_A r_B}{r_C^2} \right) \right)
\label{ppn}
\ee
where $\gamma$ is a dimensionless PPN parameter. Comparing Eq. (\ref{time-delay}) and Eq. (\ref{ppn}), we get
\be
|\gamma - 1| = \frac{GM|15\pi - 8 -3\pi\epsilon|}{2c^2r_c \ln(\frac{4 r_A r_B}{r_C^2})}~.
\label{gamma-1}
\ee
Using the measurement of the Cassini spacecraft \cite{Will:2014kxa, Bertotti:2003rm}, namely $|\gamma - 1| < 2.3 \times 10^{-5}$, and setting $r_A = 1~AB$, $r_B = 8.46~AB$ (with $AB=152.03\times 10^{9}$~m  to be the distance between the Sun and the Earth), $r_C = 1.6~R_\odot$, and $M=M_{\odot}$, we obtain
\be
-44.9 < \epsilon < 53.2~,
\label{shapiro-epsilon_1}
\ee
and in order to be consistent with the constraint $\epsilon \leq 1$, the above bounds could be modified to
\be
-44.9 < \epsilon < 1~.
\label{shapiro-epsilon_2}
\ee
\par\noindent
Assuming the ``worst'' situation, namely $\epsilon \simeq -44.9$, then Eq. (\ref{alpha_0-epsilon}) gives the upper bound for the dimensionless LQGUP parameter $\alpha_0$ to be
\be
|\alpha_0| \leq 3.17 \times 10^{38}
\label{shapiro-alpha_0}
\ee
\par\noindent
which is comparable to the bound obtained before from the observational test of the light deflection.
%
%
%
%
\subsection{Gravitational Redshift}
%
%
\par\noindent
We adopt the analysis of Ref. \cite{Okcu:2021oke} for the LQGUP-deformed Schwarzschild metric in order to calculate the change of electromagnetic signal frequency moving from the Earth surface, so $r_A = R_\oplus$, to height $h$, so $r_B = R_\oplus +h$. Therefore, we get
\be
\frac{G M ((3-\epsilon) r_A + (1-\epsilon) r_B)}{2 r_A r_B c^2} < 0.01
\label{grav-red-shift}
\ee
where we use the Pound-Snider experiment results \cite{Pound:1965zz}. Utilizing the mass of Earth $M_\oplus = 5.972 \times 10^{24}$~kg and its radius $R_\oplus = 6.378 \times 10^{6}$~m, with the height at which the experiment has been performed, namely $h = 22.86$~m, we obtain
\be
-1.4 \times 10^7 < \epsilon~.
\label{red-shift-epsilon}
\ee
\par\noindent
Employing the lower bound of Eq.(\ref{red-shift-epsilon}) in Eq.(\ref{alpha_0-epsilon}), we get the upper bound of dimensionless LQGUP parameter $\alpha_0$ to be
\be
|\alpha_0| \leq 1.72 \times 10^{33}
\label{red-shift-alpha_0}
\ee
which is much tighter bound than the one obtained using the Shapiro delay time observational test and comparable to the bound obtained from the Mercury  perihelion precession test.
%
%
%
%
\subsection{Geodetic Precession}
%
%
%
%
%
\par\noindent
We adopt the analysis of Ref. \cite{Okcu:2021oke} for the geodetic precession of the solar system for the LQGUP-deformed Schwarzschild metric, and approximately we obtain
\be
\Delta \Phi_{geodetic} = \Delta \Phi _{GR} \left( 1 - \frac{2\epsilon G M}{3 R c^2} \right)
\label{geodetic-precession}
\ee
\par\noindent
where $\Delta \Phi_{GR} = \frac{3\pi G M}{R c^2}$ is the geodetic precession theoretically predicted by GR. At an altitude of $6.42 \times 10^{5}$~m above the Sun surface, and with an orbital period of $97.65$~min, the Gravity Probe B (GPB) \cite{Everitt:2011hp} measures the geodetic precession to be $\Delta \Phi_{geodetic} = (6601.8 \pm 18.3)$~ mass/year while the GR prediction gives $\Delta \Phi_{GR} = 6601.1$~mass/year. Therefore, we set a bound on the dimensionless parameter $\epsilon$ as
\be
-5 \times 10^6 < \epsilon \leq 1
\label{geod-prec-epsilon}
\ee
\par\noindent
where we have used the mathematical constraint $\epsilon \leq 1$ with $M=M_\oplus$ and $R=R_\oplus$. Substituting the lower bound of  the dimensionless parameter $\epsilon$ given by Eq.(\ref{geod-prec-epsilon}) into  Eq.(\ref{alpha_0-epsilon}), we obtain the upper bound of the dimensionless LQGUP parameter $\alpha_0$ to be
\be
|\alpha_0| \leq 1.72 \times 10^{33}
\label{geod-prec-alpha_0}
\ee
which is exactly the same with the upper bound obtained before using the observational test of the gravitational redshift.

\section{Conclusion and Discussions}
%
%
%
%
\par\noindent
In this letter, we compared the temperature of Schwarzschild black hole when modified due to the linear and quadratic terms in the momenta of the generalized uncertainty principle (LQGUP) with the temperature obtained when the Schwarzschild black hole metric is corrected in a quantum context. The dimensionless LQGUP parameter, i.e., $\alpha_{0}$, was expressed in terms of the dimensionless parameter $\epsilon$ that characterizes the quantum-gravity corrected Schwarzschild black hole metric. By employing six solar system-based observational (gravitational) tests: light deflection, perihelion precession, pulsar periastron shift, Shapiro time delay, gravitational redshift, and geodetic precession, we utilized the bounds on the dimensionless parameter $\epsilon$ to obtain bounds on the dimensionless LQGUP parameter. In the literature, there are the corresponding bounds for the dimensionless parameter $\beta_0$ of the generalized uncertainty principle with only quadratic terms in momentum (GUP). We listed all of these bounds, together with the new bounds we derived in this work, in \hyperref[tab1]{the table} below. In order to compare them, one can roughly assume that $\beta_{0}\sim\alpha_{0}^{2}$. It is evident that the bounds based on LQGUP are much tighter than the corresponding bounds based on GUP. 
\par\noindent
Finally, it should be stressed that, all these bounds are much weaker than the bound set by the electroweak scale $\ell_{EW} \le 10^{17} \ell_{Planck}$ as well as weaker than the bounds set by other gravitational tests performed in strong gravitational fields. One can say that this is anticipated since all the bounds in this work are obtained by employing solar system-based gravitational tests and in the solar system the QG effects are not so strong to be detected as in other gravitational systems such as that of the two stellar mass black holes that are merging and produce a spacetime of high curvature.  In the latter case, one can detect the emitted gravitational waves and set strict bounds on the dimensionless LQGUP parameter, for instance from the events GW150914 and GW170814 one can get $\alpha_{0}\sim 10^{8}$ \cite{Das:2021lrb}. Of course, one can study pure quantum systems to measure quantum effects and, thus, to get strict bounds on the dimensionless LQGUP parameter, for instance from the Lamb shift one gets $\alpha_{0}\sim 10^{10}$ \cite{Ali:2011fa}.

\begin{center}\label{tab1}
\begin{tabular}{ l@{\hskip  0.3in}c@{\hskip 0.3in}c }
\hline
Experiment & $\beta_0$ (GUP) &  $\alpha_0$ (LQGUP) \\ 
\hline \\
Light deflection & $5.3 \times 10^{78}$ \cite{Scardigli:2014qka} & $3.50\times 10^{38}$ [NEW]\\
Perihelion precession & $3.0\times 10^{72}$ \cite{Scardigli:2014qka} & $1.38\times 10^{33}$ [NEW]\\
Pulsar Periastron shift & $2.0\times10^{71}$ \cite{Scardigli:2014qka} & $2.96\times10^{31}$ [NEW]\\
Shapiro Time Delay & $3.6 \times 10^{78}$ \cite{Okcu:2021oke} & $3.17 \times 10^{38}$ [NEW]\\
Gravitational Redshift & $1.1 \times 10^{73}$ \cite{Okcu:2021oke} & $1.72 \times 10^{33}$ [NEW]\\
Geodetic Precession & $3.7 \times 10^{72}$ \cite{Okcu:2021oke}  & $1.72 \times 10^{33}$ [NEW]\\  
\\
\hline
\end{tabular}
\end{center}

\pagebreak

\end{arabicfootnotes}

\begin{thebibliography}{99}
%
%
%
%
%
%
\bibitem{polchinski_1}
J.~Polchinski, String theory. Vol. \textbf{1}: An introduction to the bosonic string, .
Cambridge, UK: Univ. Pr. (1998).


\bibitem{polchinski_2}
J. Polchinski, String theory. Vol. \textbf{2}: Superstring theory and beyond, . Cambridge, UK: Univ. Pr. (1998).

\bibitem{Smolin:2004sx}
L.~Smolin,
[arXiv:hep-th/0408048 [hep-th]].

\bibitem{Rovelli:1990pi}
C.~Rovelli,
Class. Quant. Grav. \textbf{8}, 317-332 (1991)

\bibitem{Loll:2019rdj}
R.~Loll,
Class. Quant. Grav. \textbf{37}, no.1, 013002 (2020)
[arXiv:1905.08669 [hep-th]].



\bibitem{Amati:1987wq}
D.~Amati, M.~Ciafaloni and G.~Veneziano,
Phys. Lett. B \textbf{197}, 81 (1987)


\bibitem{Gross:1987kza}
D.~J.~Gross and P.~F.~Mende,
Phys. Lett. B \textbf{197}, 129-134 (1987)
  
  
  
\bibitem{Amati:1988tn}
D.~Amati, M.~Ciafaloni and G.~Veneziano,
Phys. Lett. B \textbf{216}, 41-47 (1989)
  
\bibitem{Konishi:1989wk}
K.~Konishi, G.~Paffuti and P.~Provero,
Phys. Lett. B \textbf{234}, 276-284 (1990)
  
  
\bibitem{Veneziano:1990zh}
G.~Veneziano,
CERN-TH-5889-90.


\bibitem{Brau:1999uv}
F.~Brau,
J. Phys. A \textbf{32}, 7691-7696 (1999)
[arXiv:quant-ph/9905033 [quant-ph]].


\bibitem{Das:2008kaa}
S.~Das and E.~C.~Vagenas,
Phys. Rev. Lett. \textbf{101}, 221301 (2008)
[arXiv:0810.5333 [hep-th]].

\bibitem{Das:2010sj}
S.~Das and E.~C.~Vagenas,
Phys. Rev. Lett. \textbf{104}, 119002 (2010)
[arXiv:1003.3208 [hep-th]].

\bibitem{Pedram:2011xj}
P.~Pedram, K.~Nozari and S.~H.~Taheri,
JHEP \textbf{03}, 093 (2011)
[arXiv:1103.1015 [hep-th]].


\bibitem{Das:2009hs}
S.~Das and E.~C.~Vagenas,
Can. J. Phys. \textbf{87}, 233-240 (2009)
[arXiv:0901.1768 [hep-th]].


\bibitem{Ali:2009zq}
A.~F.~Ali, S.~Das and E.~C.~Vagenas,
Phys. Lett. B \textbf{678}, 497-499 (2009)
[arXiv:0906.5396 [hep-th]].


\bibitem{Ali:2010yn}
A.~F.~Ali, S.~Das and E.~C.~Vagenas,
[arXiv:1001.2642 [hep-th]].


\bibitem{Vagenas:2018pez}
E.~C.~Vagenas, A.~Farag Ali and H.~Alshal,
Eur. Phys. J. C \textbf{79}, no.3, 276 (2019)
[arXiv:1811.06614 [gr-qc]].


\bibitem{Vagenas:2019rai}
E.~C.~Vagenas, A.~F.~Ali and H.~Alshal,
Phys. Rev. D \textbf{99}, no.8, 084013 (2019)
[arXiv:1903.09634 [hep-th]].


\bibitem{Vagenas:2019wzd}
E.~C.~Vagenas, A.~F.~Ali, M.~Hemeda and H.~Alshal,
Eur. Phys. J. C \textbf{79}, no.5, 398 (2019)
[arXiv:1903.08494 [hep-th]].


\bibitem{Vagenas:2020bys}
E.~C.~Vagenas, A.~Farag Ali, M.~Hemeda and H.~Alshal,
Annals Phys. \textbf{432}, 168574 (2021)
[arXiv:2008.09853 [hep-th]].

\bibitem{Robertson:1929zz}
H.~P.~Robertson,
Phys. Rev. \textbf{34}, 163-164 (1929)
doi:10.1103/PhysRev.34.163


\bibitem{Vagenas:2017vsw}
E.~C.~Vagenas, S.~M.~Alsaleh and A.~Farag,
EPL \textbf{120}, no.4, 40001 (2017)
[arXiv:1801.03670 [hep-th]].


\bibitem{FaragAli:2015boi}
A.~F.~Ali, M.~M.~Khalil and E.~C.~Vagenas,
EPL \textbf{112}, no.2, 20005 (2015)
[arXiv:1510.06365 [gr-qc]].

\bibitem{Donoghue:1993eb}
J.~F.~Donoghue,
Phys. Rev. Lett. \textbf{72}, 2996-2999 (1994)
[arXiv:gr-qc/9310024 [gr-qc]].


\bibitem{Donoghue:1994dn}
J.~F.~Donoghue,
Phys. Rev. D \textbf{50}, 3874-3888 (1994)
[arXiv:gr-qc/9405057 [gr-qc]].

\bibitem{Bjerrum-Bohr:2002gqz}
N.~E.~J.~Bjerrum-Bohr, J.~F.~Donoghue and B.~R.~Holstein,
Phys. Rev. D \textbf{67}, 084033 (2003)
[erratum: Phys. Rev. D \textbf{71}, 069903 (2005)]
[arXiv:hep-th/0211072 [hep-th]].


\bibitem{Scardigli:2016pjs}
F.~Scardigli, G.~Lambiase and E.~Vagenas,
Phys. Lett. B \textbf{767}, 242-246 (2017)
[arXiv:1611.01469 [hep-th]].


\bibitem{Scardigli:2014qka}
F.~Scardigli and R.~Casadio,
Eur. Phys. J. C \textbf{75}, no.9, 425 (2015)
[arXiv:1407.0113 [hep-th]].


\bibitem{Okcu:2021oke}
\"O.~\"Okc\"u and E.~Aydiner,
Nucl. Phys. B \textbf{964} (2021), 115324
[arXiv:2101.09524 [gr-qc]].


\bibitem{Pound:1965zz}
R.~V.~Pound and J.~L.~Snider,
Phys. Rev. \textbf{140}, B788-B803 (1965)



\bibitem{Das:2021lrb}
A.~Das, S.~Das, N.~R.~Mansour and E.~C.~Vagenas,
Phys. Lett. B \textbf{819}, 136429 (2021)
[arXiv:2101.03746 [gr-qc]].


\bibitem{Ali:2011fa}
A.~F.~Ali, S.~Das and E.~C.~Vagenas,
Phys. Rev. D \textbf{84}, 044013 (2011)
[arXiv:1107.3164 [hep-th]].

\bibitem{Will:2014kxa}
C.~M.~Will,
Living Rev. Rel. \textbf{17}, 4 (2014)
doi:10.12942/lrr-2014-4
[arXiv:1403.7377 [gr-qc]].

\bibitem{Bertotti:2003rm}
B.~Bertotti, L.~Iess and P.~Tortora,
Nature \textbf{425}, 374-376 (2003)
doi:10.1038/nature01997

\bibitem{Shapiro:2004zz}
S.~S.~Shapiro, J.~L.~Davis, D.~E.~Lebach and J.~S.~Gregory,
Phys. Rev. Lett. \textbf{92}, 121101 (2004)
doi:10.1103/PhysRevLett.92.121101

\bibitem{Lambert:2009xy}
S.~B.~Lambert and C.~Le Poncin-Lafitte,
Astron. Astrophys. \textbf{499}, 331 (2009)
doi:10.1051/0004-6361/200911714
[arXiv:0903.1615 [gr-qc]].

\bibitem{Weisberg:2010zz}
J.~M.~Weisberg, D.~J.~Nice and J.~H.~Taylor,
Astrophys. J. \textbf{722}, 1030-1034 (2010)
doi:10.1088/0004-637X/722/2/1030
[arXiv:1011.0718 [astro-ph.GA]].

\bibitem{Shapiro:1964uw}
I.~I.~Shapiro,
Phys. Rev. Lett. \textbf{13}, 789-791 (1964)
doi:10.1103/PhysRevLett.13.789

\bibitem{Everitt:2011hp}
C.~W.~F.~Everitt, D.~B.~DeBra, B.~W.~Parkinson, J.~P.~Turneaure, J.~W.~Conklin, M.~I.~Heifetz, G.~M.~Keiser, A.~S.~Silbergleit, T.~Holmes and J.~Kolodziejczak, \textit{et al.}
Phys. Rev. Lett. \textbf{106}, 221101 (2011)
doi:10.1103/PhysRevLett.106.221101
[arXiv:1105.3456 [gr-qc]].






\end{thebibliography}
\end{document}